\documentclass[12pt]{iopart}

\usepackage{graphicx}  
\usepackage{color}
\usepackage{amstext}
\usepackage{calc}
\begin{document}

\title[Gender differences in understanding of Newtonian mechanics]{Gender differences in conceptual understanding of Newtonian mechanics: a UK cross-institution comparison}

\author{Simon Bates,\footnote{Current address: Department of Physics and Astronomy, University of British Columbia, Vancouver, V6T 1Z1, Canada} Robyn Donnelly, Cait MacPhee}

\address{School of Physics and Astronomy, University of Edinburgh, Edinburgh, EH9 3JZ, UK} 

\ead{r.c.a.donnelly@ed.ac.uk}

\author{David Sands}

\address{Department of Physics and Mathematics, University of Hull, Kingston-upon-Hull, HU6 7RX, UK}

\author{Marion Birch, Niels R. Walet}

\address{School of Physics and Astronomy, University of Manchester, Manchester, M13 9PL, UK}

\begin{abstract}
  We present results of a combined study from three UK universities
  where we investigate the existence and persistence of a performance
  gender gap in conceptual understanding of Newtonian mechanics. Using
  the Force Concept Inventory, we find that students at all three
  universities exhibit a statistically significant gender gap, with
  males outperforming females. This gap is narrowed but not eliminated
  after instruction, using a variety of instructional
  approaches. Furthermore, we find that before instruction the
  quartile with the lowest performance on the diagnostic instrument
  comprises a disproportionately high fraction ($\sim$ 50\%) of the total
  female cohort. The majority of these students remain in the
  lowest-performing quartile post-instruction. Analysis of responses
  to individual items shows that male students outperform female
  students on practically all items on the instrument. Comparing the
  performance of the same group of students on end-of-course
  examinations, we find no statistically significant gender gaps.
\end{abstract}

\submitto{\EJP}

\maketitle

\section{Introduction}

There are ongoing concerns about female participation and performance in academic
physics.  Here we examine the gender differences in performance in introductory Newtonian
mechanics within introductory physics courses at three UK
universities, using the Force Concept Inventory (FCI), a standard
diagnostic test of conceptual understanding. Developed by Hestenes and
co-workers at Arizona State in the early 1990's, the FCI has become
the gold standard diagnostic test of conceptual understanding in
physics \cite{FCI}. Following Hake's landmark paper in 1998
\cite{Hake}, in which the efficacy of different instructional
methodologies was evaluated using this instrument with a sample of
nearly six thousand students, it has become widely accepted and used
within the discipline community. It has also served as the benchmark
for the subsequent creation of a wide variety of instruments and
inventories for use in science teaching and learning, currently
numbering over 50 \cite{EdWebSTEM}. As an estimate, since being
developed the FCI has likely been deployed to hundreds and thousands of
students worldwide to date.

Several studies from US institutions have previously investigated the
performance gender gap in introductory physics courses using the FCI
instrument. (We use gender in this context as equivalent to sex; we
are aware that this can rightfully attract criticism \cite{gender}.)
The standard methodology of implementation is to assess students with
the instrument prior to any instruction on the course and then again
after instruction (the so-called `pre- and post-' test methodology).
Lorenzo {\it et al} \cite{CrouchMazur} presented extensive data from
Harvard students undertaking an introductory calculus based course
between 1990 and 1997. Their data indicate the presence of a
statistically-significant gender gap on the pre-instruction tests,
expressed as difference in the mean score by male students and the
mean score of female students.  Their findings show male students
consistently outperforming female students. Furthermore, they contend
that certain instructional methodologies are more effective at
reducing (or eliminating) the performance gender gap, specifically the
interactive engagement style of lecture instruction emphasising peer
instruction and discussion that Eric Mazur developed in the late 1990s
in response to students' initial performance on the FCI at Harvard. In
a replication study at the University of Colorado, Pollock {\it et al}
\cite{PRSTPER2007, Noah} found that use of interactive engagement
instructional methodologies is not necessarily sufficient to close or
eliminate the performance gender gap on the FCI, citing examples
where, despite overall improvements in performance between pre- and
post-instruction tests, the gap was found to widen. They suggest there
are additional effects due to differences in student preparation and background, as well variability in instructors. 
Docktor and Heller \cite{DocktorHeller} have
presented an analysis of data collected from 40 separate classes, 5500
students and 22 different instructors. They too confirm the presence
of a significant gender gap pre-instruction, which persists
post-instruction overall, though the individual (per class) changes in
the performance gender gap span a broad range from +7\% (gap widens)
to -6\% (gap narrows).

We are motivated to re-examine this in the context of UK physics
undergraduates, in part because of the lack of clear consensus in the
literature, but moreover because both the style of university education
and the preparation prior coming to university are very different from
those in the USA. Additionally, many of the studies referred to above 
relate to courses delivered to non-majors in Physics, sometimes with 
atypical gender profiles (for example, the Harvard courses referred to 
above have a male:female student ratio of less than 2, whereas undergraduate
courses in the UK typically have ratios closer to 4). Our previous studies, using a
widely-accepted instrument to survey students' attitudes and beliefs
about physics, have indicated there can be significant differences in
the response profiles of US and UK students undertaking introductory
physics courses \cite{BatesPRSTPER}, and this is worthy of further
investigation with respect to gender and performance on the FCI.

We apply our investigation to a typical UK cohort of students, in which female
students are under-represented. Over the past 10 years of A-level 
examinations (the most common school leaving examination in England, 
Wales and Northern Ireland) we have seen a dramatic rise in the 
overall uptake of mathematics, a slight decline followed by a slight
rise in physics, and a minimal rise in chemistry. Over the same time
period the proportion of females taking physics has dropped
marginally, from 23 to 21\%, and is much lower than comparable 
figures for mathematics (constant at about 40\%) and chemistry 
(a slight decrease from 51 to 47\%) \cite{RDFYP3}. There is at 
least some evidence that female students taking a physics A-level 
exam are more able than their male colleagues, but at the same time 
their motivation for studying physics may be much more geared towards the medical sciences \cite{Stewart}.

At tertiary level, the Institute of Physics (IoP) has collected data that
shows that the proportion of females entering first year undergraduate physics in the UK
has remained between 18-20\% for the last 15 years \cite{RDFYP6}.
Female participation is, therefore, a key issue. Allied to that is the
issue of retention as students progress through their education.  The
relatively poor progression of females to the most advanced courses in
physics has been referred to as a `leaky pipeline' \cite{Kate}, with
proportionately more female students leaving programmes of study at
early stages. It is a contributing factor to the fact that, based on US data, physics has one of the
lowest proportion of female PhD graduates of all disciplines 
\cite{graph}.

This study presents results from pre- and post-instruction testing
using the FCI at three different UK universities: Edinburgh, Hull and
Manchester. Our study has the following aims:
\begin{enumerate}
\item To evaluate the existence and possible persistence of a
  performance gender gap in introductory (first year of study)
  physics courses;
\item To evaluate individual test items that display significant
  differences in performance of male / female students.
\end{enumerate}

This paper is organised as follows: in the next section we detail the
different course contexts at the three institutions and the
methodology used to obtain and analyse test data. We then present
results from the three universities and in the final section discuss
both these results and their implications.

\section{Methodology}

\subsection{UK education context}

We briefly summarise the educational background from which the vast
majority of our undergraduate intake is drawn for readers not familiar
with the UK system. There are three main sets of qualifications for our
intakes:

\begin{itemize}
\item A-levels, taken by the majority of students from England, Wales
  and Northern Ireland, and some from Scotland.  The final two A level
  years follow on from six years of primary schooling and five years
  at secondary school where a broad set of subjects is studied.  Students
   generally take a minimum of three subjects for these final two years
  before going to university. The results for each subject are
  reported as letter grades (highest grade A*, other pass grades A-E,
  and U=``unclassified'' as a failing grade) and are obtained from a
  combination of the results from a set of modular exams.  Within this
  system there are various syllabuses for both mathematics and
  physics, and physics itself is taught in manner which largely avoids
  the use of calculus. Students are exposed to more of the
  mathematical rigour of physics if they choose so-called ``mechanics
  modules" (elective modules on the mathematics of Newtonian
  Mechanics) as part of their mathematics A-level.
\item Scottish Higher qualifications, taken by most Scottish students,
  since Scotland has a separate education system. Entry to
  universities in Scotland is based on performance in these
  qualifications, taken at the end of year five of secondary
  level. Typically 6 subjects are studied to Higher level. Most
  students stay on for a sixth year at secondary level, taking
  (usually) three subjects at Advanced Higher level.
\item Finally, a growing fraction of our students enter with the UK
  International Baccalaureate. It consists of three specific core
  elements, and study of six elective subjects, three of
  which are at higher level.
\end{itemize}

The typical age of an incoming student to undergraduate physics
programmes in the UK is 18 or 19 (occasionally 17 from Scotland),
depending on their date of birth. The school system in the UK is
funded through a variety of sources, where the most important
difference with the rest of Europe is the larger number of students
studying at privately funded schools.  First year students at the
three universities in this study (two from England, one from Scotland)
therefore comprise a diverse group in terms of their prior
academic background and ability.
 
\subsection{Institutional contexts and course details}

All three institutions have been utilising the FCI (or a close variant
thereof) as an assessment instrument within their introductory physics
courses for a number of years. We all had slight differences in our
approach to the FCI, but we have aligned our processes in the 2011-12
academic year, on which this work is based.
Due to a substantial increase in tuition fees in England for 2012, this
was also a year when entry qualifications were higher than in 
recent previous years.

All three universities require both physics and maths school-leaving
qualifications for students wanting to study physics
courses. There are, naturally, differences between the cohorts and
courses at the three universities, which are detailed below.

\subsubsection{Edinburgh}

The School of Physics and Astronomy at the University of Edinburgh has
an annual undergraduate intake onto the physics degree programme of about 
120 students, of whom 21-25\% are female. Approximately 60\% of students 
enter the University with Scottish school-leaving qualifications (Highers 
or Advanced Highers), 30\% have taken A-level exams, and the remainder 
have taken other qualifications such as the International Baccalaureate. 

The Scottish Bachelor's degree has a normal duration of
four years (one year longer than in the rest of the UK), with a first
year that is slightly broader than that in England. The first
year class studied here comprises students for whom physics is a mandatory requirement
for their degree programme (mainly students on physics degrees) and
those who are taking it as an outside subject or an elective. In
recent years, each of these constituents comprised about half the
class, thus the total class size ranges from 200 to 300 students each
year. 

The eleven-week course has for many years been a focal point for curriculum
innovation within the School, and details of the instructional design
\cite{(ref ALTC2006)}, the role of studio-based workshop classes
\cite{ref CAL 2006}, student generated assessment content \cite{(PERC
  2011)} and the move to `invert' the traditional lecture environment
\cite{Ed ref HEA 2012} have been reported elsewhere. In the 2011-12
presentation of the course, for which we report data here, the most
significant change was the inclusion of the latter two interventions
(student-generated assessment content and the `inverted' classroom
approach) in the standard presentation of the course.

\subsubsection{Hull}

At the University of Hull, the first year physics intake has doubled
in the last four years and stood at 70 students in 2011/12, 10\% of
whom were female. The vast majority of students enter the 3-year BSc programme 
with A-level qualifications. 

Classical mechanics is taught through a modelling
curriculum \cite{modelling}.  As is common in modelling instruction, we base
student-generated models on group discussion, but due to institutional
constraints the course was delivered in a conventional lecture
theatre, which meant that, at best, the discussion was limited to
neighbouring pairs of students. The course is taught over a ten-week
period in the first semester of the first year, with each lecture
mixing elements of formal instruction, interactive engagement and
discussion between neighbours. The formal instruction is based on a
structured approach to the use of multiple representations in
constructing models, with the role of representations in evaluating,
describing, analysing and solving problems being
emphasized. Interactive engagement and peer discussion are used
primarily to provide students with opportunities to use multiple
representations for themselves. Details of the method and the
conceptual gains have been described elsewhere \cite{David}.

\subsubsection{Manchester}
The first year undergraduate intake to the School of Physics and
Astronomy at the University of Manchester comprises between 230 and
290 students per year, of which approximately 20\% are females. Nearly
all the students are registered on either a 4 year MPhys or a 3 year
BSc degree in Physics, some with a subsidiary subject such as
Astrophysics. A small fraction of the students are registered on a
joint Mathematics and Physics degree programme.

The vast majority of entrants possess A-level qualifications, however 
the students' prior experience of Newtonian mechanics varies considerably 
depending on their choices of optional modules within their Mathematics 
A level and, if they chose
to do take it, Further Mathematics A level. The median number of
Mechanics modules taken is 2, but some students do not study any of
these optional mechanics modules at A level whereas others may have
studied up to four or five (depending on which of the five A-level
examination board papers they have sat).

All students take a Newtonian mechanics course in their first semester
at university. The eleven-week course has been taught in a
non-traditional manner for several years, using interactive techniques
such as an electronic voting system (`clickers'), peer instruction \cite{Mazur} and
Just-in-Time Teaching \cite{JITT}. A comprehensive suite of e-learning material is
used to support the students' learning and weekly online assignments
encourage students to consolidate their understanding as the course
progresses. These teaching techniques have had a positive impact
compared with the previous traditional approach, both in terms of
student satisfaction and examination
performance \cite{BirchWaletPubs, BW2}.

\subsection{Implementation of the FCI}

A summary of the implementation procedures employed at the three
institutions is presented in Table~\ref{tab:FCI}. In the case of Edinburgh, where the FCI score contributed to
approximately 3\% to the course mark, it was the better of a student's
two attempts that counted.

\begin{table}
\caption{\label{tab:FCI}FCI implementation details at participating institutions}
\begin{indented}
\item[]\begin{tabular}{llllll}
\br
\textrm{University}&
\textrm{FCI used }&
\textrm{Delivery}&
\textrm{Time}&
\textrm{Timing}&
\textrm{Contibution to}\\

\textrm{}&
\textrm{since}&
\textrm{mechanism}&
\textrm{limit}&
\textrm{pre- / post-}&
\textrm{final mark (\%)}\\
\br

Edinburgh & 2006$^{\rm a}$  & Online & 90 mins & weeks 1 /  8 & 3 \\
Hull & 2008$^{\rm b}$ & Paper & none & weeks 0 /10 & 0 \\
Manchester & 2008 & Paper & 60 mins & weeks 0 / 6 & 0\\
\br 
\end{tabular}
\item[] $^{\rm a}$ Between 2006 and 2010 a variant of the FCI was used with additional questions.
\item[] $^{\rm b}$ Matched pre- and post-data were collected for the first time in 2011.
\end{indented}
\end{table}

\subsection{Statistical tests}

Since our data is represented in two two-way contingency tables (one
dimension is male or female, the other their score on one of the FCI
tests), the best approach to see whether the differences between males
and female students are significant is the Pearson $\chi^2$ test of
association. This test is based on a test statistic that measures the
divergence of the observed data from the values that would be expected
under the null hypothesis of no association. This requires calculation
of the expected values based on the data; the expected value for each
cell $(i,j)$ in a two-way table $d_{ij}$ is equal to
\[{\bar d}_{ij}=(\sum_{i'} d_{i'j})(\sum_{j'} d_{ij'})/n\] where $n$
is the total number of observations in the table. The statistic is
defined as
\[\chi^2=\sum_{ij}(d_{ij}-{\bar d}_{ij})^2/{\bar d}_{ij} .\] This is
distributed as a $\chi^2$-distribution with
$(N_{column}-1)(N_{row}-1)$ degrees of freedom, and we can test
whether the observed value for the statistic is significant by looking
at the $P$ value for the $\chi^2$ distribution.

The alternative of approximating the two sets of mark distribution as
continuous is also attractive; the disadvantage of that approach is
that the data are not normally distributed, mainly due to the fact that 
scores on the post-instruction test 
are close to the maximum, and thus fluctuations above the
mean have only a limited range. Thus the standard $t$-test for the
equality of the means does not apply. We have applied the Mann-Whitney U test to test the difference between the distributions of the independent data sets.

\section{Results} 

\subsection{Quantitative data analysis}

The FCI was administered to each cohort before and after relevant
instruction. In the presentation of results that follows, only matched
pairs of data (i.e. data for students who had taken {\it both} the
pre- and post-instruction tests) are included. Thus the sample size for each
institution is lower than total class size. 

For the members of each cohort undertaking both a pre- and
post-instruction test, we calculate the pre- and post-instruction
average percentage scores $\langle
x\rangle _{\text{pre}}$ and $\langle x\rangle _{\text{post}}$. This
allows us to
calculate a cohort-averaged normalised gain, $\langle g\rangle $ defined as
\begin{equation}
\langle g\rangle  = \frac{\langle x\rangle _{\text{post}}-\langle x\rangle _{\text{pre}}} {100-\langle x\rangle _{\text{pre}}  }
\end{equation}
This normalised gain is often considered as a measure of instructional
effectiveness, representing the fractional improvement in
understanding, as first described in Hake's study \cite{Hake}. The
normalised gain can be calculated for the entire cohort, or just the
male or female sub-cohorts, using the appropriate mean FCI scores.

For the analysis of the male and female sub-cohorts, we can define a
performance gender gap, $G$, as the difference between male and female
mean scores, such that
\begin{equation}
 G_i = \langle x\rangle ^{\text{male}}_{i} - \langle x\rangle ^{\text{female}}_{i}
\end{equation}
where the subscript $i$ denotes pre- or post-instruction, and the
convention we have adopted is that {\it positive} gaps imply male
students outperforming female and vice versa. We can define a change
in this gap, $\Delta G$, as the difference between values of $G$
determined for post- and pre-instruction, such that
\begin{equation}
 \Delta G = G_{\text{post}} - G_{\text{pre}}
\end{equation}
where the convention here is that a positive value of $\Delta G$
denotes a gap that widens as a result of instruction, and  negative 
$\Delta G$ denotes a gap that narrows.

\begin{figure}
\begin{center}
\includegraphics[width=3.5 in]{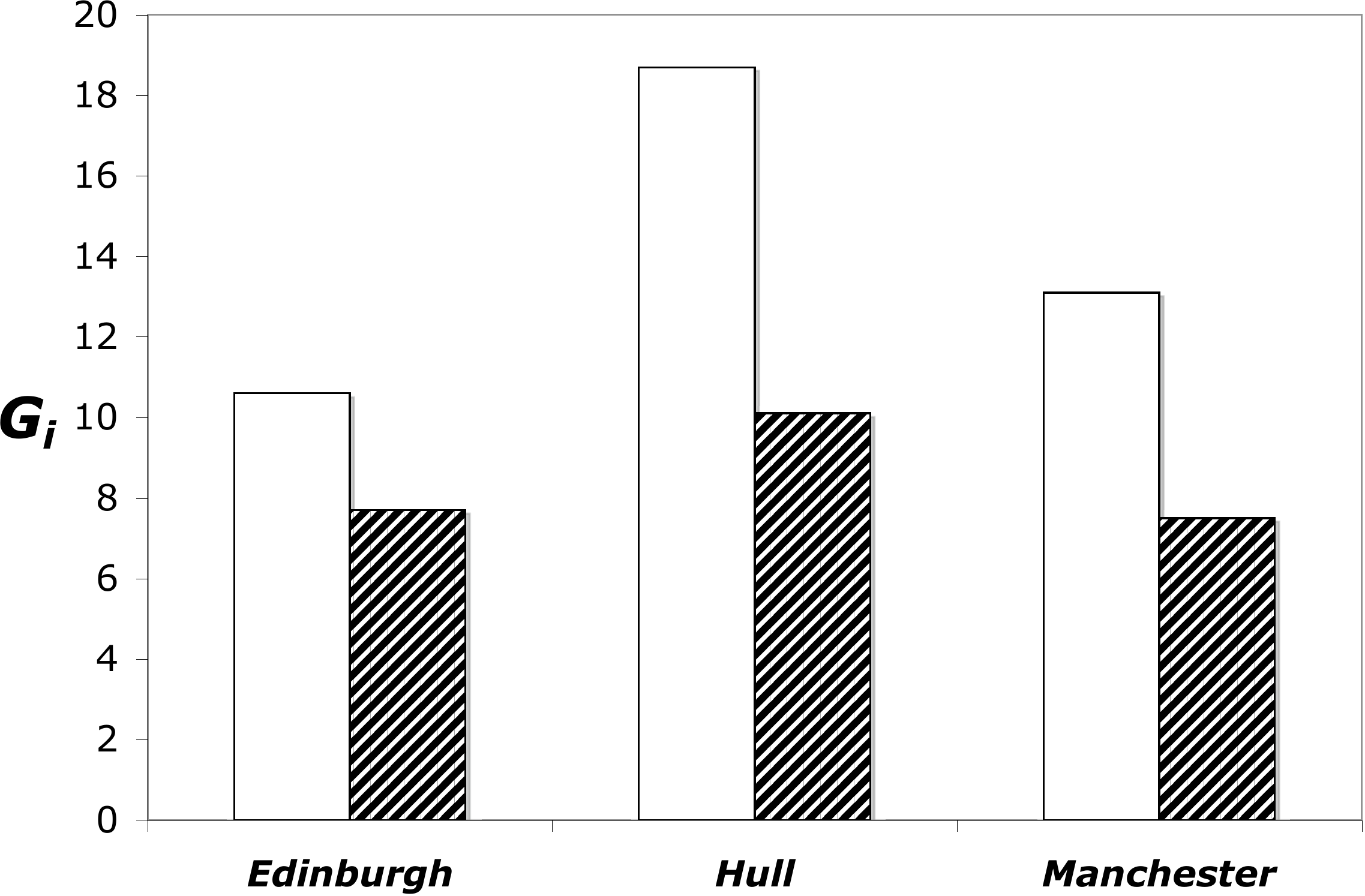}
\end{center}
\caption{\label{fig:gainmf}Gender gap, $G_i$, defined as the
  difference between the mean male cohort score and the mean female
  cohort score on the FCI, for pre-instruction (white) and
  post-instruction (hatched) testing}
\end{figure}

Table~\ref{tab:rawdata} presents values for these quantities for all
three institutions. Cohorts from all three institutions show
substantial learning gains on the FCI, comparable with those seen on
`reformed' courses in studies reported in the literature \cite{Hake},
providing evidence for effective (even though they are all rather
different) instructional methodologies. However, all three
institutions show a consistent performance gender gap ($G$ positive)
on the basis of pre-instruction test results, ranging from $+10\%$ to
$+19\%$. Furthermore, this gap persists on the post-instruction
assessment and is statistically significant ($p < 0.05$), but is reduced
in all three cases.  Figure \ref{fig:gainmf} illustrates this,
presenting the male and female sub-cohort data for pre- and
post-instruction tests in graphical form.

\begin{figure}
\begin{center}
\includegraphics[width=3in]{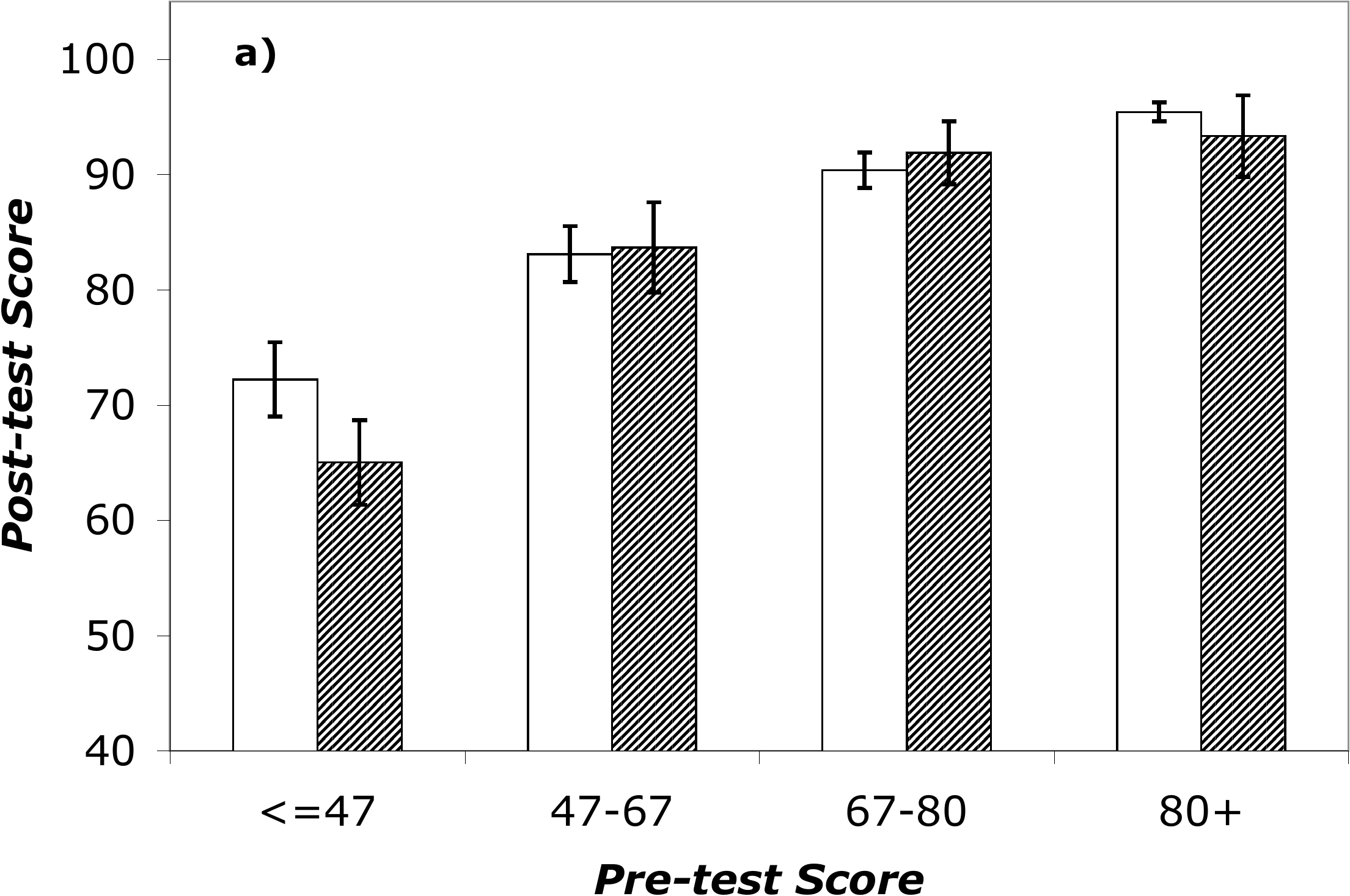}\\[0.1cm]
\includegraphics[width=3in]{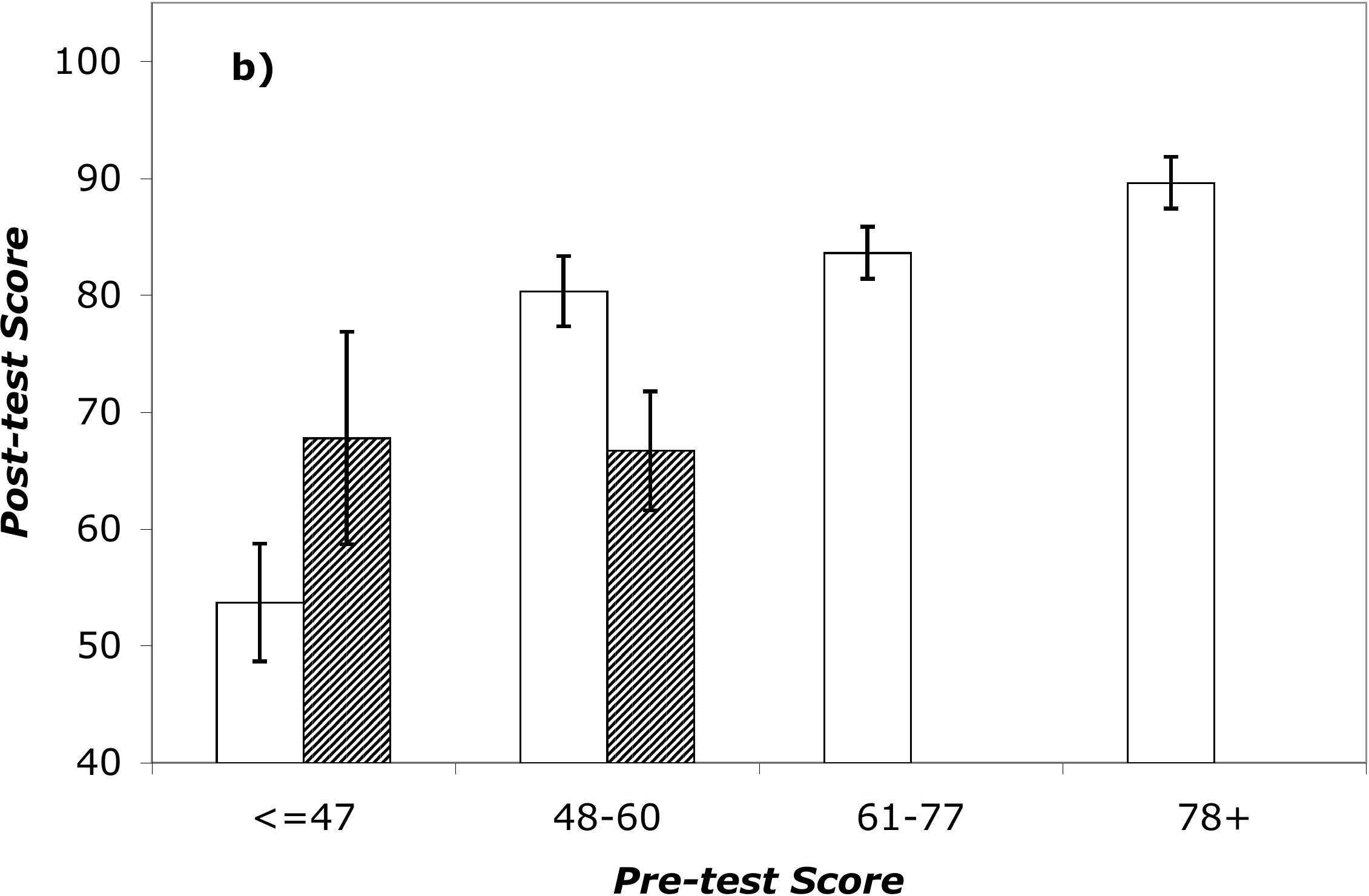}\\[0.1cm]
\includegraphics[width=3in]{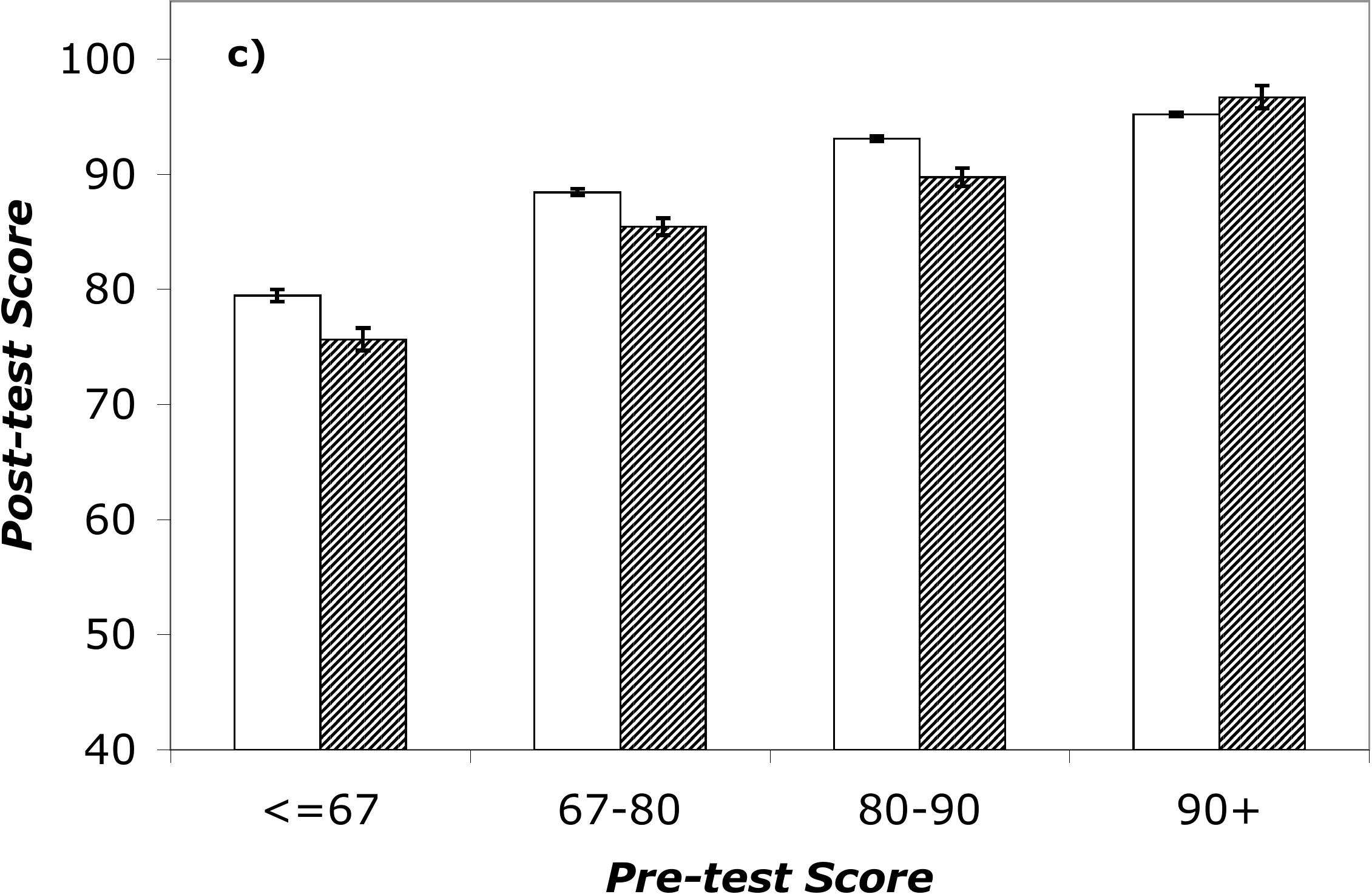}
\end{center}
\caption{\label{fig:quartiles}FCI post-instruction mean score for gender-split quartile groups. Male students are represented by white bars, female by grey. Data refers to students from (a) Edinburgh; (b) Hull; (c) Manchester. Error bars represent the standard error in the mean.}
\end{figure}

Table~\ref{tab:rawdata} illustrates that on the basis of the
pre-instruction test, female students start the courses with lower FCI
attainment. It is instructive to investigate the distribution of these
students across the cohort and chart their later outcomes on the
post-instruction test. To do this, we split the cohort on the basis of
pre-instruction test performance into quartiles (of approximately
equal size) at each institution. We then further separate each
quartile into male and female subgroups. The performance on the
post-instruction test of these gender-split quartile groups for each
institution is presented on Figure~\ref{fig:quartiles}.

\begin{table}
\caption{\label{tab:rawdata}Cohort performance on the FCI. Values in parentheses are the standard error of the mean;  see text for definition of other quantities.}
\begin{tabular}{lllcccccl}
\br
Institution & Group & $N$ & Assessment  & $\langle x\rangle $ & $\langle g\rangle$ & $G$ & $p$ &  $\Delta G$ \\ 
\mr
Edinburgh & Whole class & 161 & Pre & 64.4 (1.7) &  &  & \\
 &  & & Post& 83.9 (1.2) & 0.55  &  & \\

Hull & Whole class & 46 & Pre& 59.1 (2.7) & & & \\
& & & Post & 75.9 (2.4) &  0.41  & & \\

Manchester & Whole class & 258 & Pre& 76.4 (1.0) & & & \\
& & & Post & 87.6 (0.7) & 0.48 & & \\
\mr

Edinburgh & Male & 116 & Pre & 67.4 (1.9) & & & & \\
& Female & 45 & Pre & 56.8 (3.2) & & 10.6 & 0.005 &\\

& Male & 116 & Post & 86.0 (1.3) & 0.57 & & & \\
& Female & 45 & Post & 78.3 (2.7) & 0.50 & 7.7 & 0.013 & -2.8 \\

\mr

Hull & Male & 40 & Pre & 61.5 (2.7) & & & & \\
& Female & 6 & Pre & 42.8 (6.1) & & 18.7 & \textless 0.001 & \\

& Male & 40 & Post & 77.3 (2.6) & 0.41 & & & \\
& Female & 6 & Post & 67.2 (4.6) & 0.43 & 10.1 &\textless 0.001 & -8.6 \\

\br

Manchester & Male & 198 & Pre & 79.4 (1.0) & & & & \\
& Female & 60 & Pre & 66.3 (2.4) & & 13.1 & 0.015 &\\

& Male & 198 & Post & 89.4 (0.7) & 0.49 & & & \\
& Female & 60 & Post & 81.9 (2.0) & 0.46 & 7.5 & 0.050 & -5.6 \\

\br

\end{tabular}

\end{table}

For the data from Hull, there are no female students in the top two
quartiles and the rather small sample sizes (particularly of female
students) means it is not sensible to try and draw too many
conclusions from the data in Figure~\ref{fig:quartiles}b. For the
larger sample sizes of the Manchester and Edinburgh cohorts, there 
was no statistically significant difference in post-instruction test 
scores of the male and female cohorts within each quartile group.

However, we do note that in case of the lowest quartile
for all three institutions, the mean post-instruction test score for
this quartile barely reaches the pre-instruction whole-cohort average
for that institution. In other words, on average, students in the
lowest quartile pre-instruction show the lowest normalised gains post
instruction. Another way to analyse this data  is to consider the `churn'
between students in the lowest ability quartile on both the
pre-instruction and post-instruction tests. Considering the 
Edinburgh data, we find that approximately 70\% of students initially in the lowest
quartile are also found in the lowest quartile on the post-instruction
test, with all of the remainder elevated to just the third quartile.
For Manchester almost 60\% of those students initially in the lowest 
quartile on the basis of pre-instruction FCI scores remain there.

Further insight into the consequences of these results can be gained
if we consider the relative proportions of male and female students in
each of the four pre-instruction test quartiles: so as not to
overcrowd Figure~\ref{fig:quartiles}, this is presented separately in
Table~\ref{tab:quartiles}.

\begin{table}
\caption{\label{tab:quartiles}
Fraction of male and female students in each quartile group of pre-instruction FCI scores. $N_{\text{{tot}/M/F}} $ represents the number of students in total, those who are male and those who are female, respectively. $f_{\text{M/F}}$ gives the fraction of male or female students, respectively, who are in each of the four quartile groups Q1 (highest) to Q4 (lowest) expressed as a fraction of the total number of male or female students in the cohort.}
\begin{indent}
\begin{tabular}{@{}lccccccccccc}
\br
Institution & $N_{\text{tot}}$ & $N_{\text M}$ & $N_{\text F}$ & \multicolumn{2}{c}{\text{Q1}} &  \multicolumn{2}{c}{\text{Q2}}    &  \multicolumn{2}{c}{\text{Q3}} &  \multicolumn{2}{c}{\text{Q4}}\\
& &  & & $f_{\text M}$ & $f_{\text F}$      & $f_{\text M}$ & $f_{\text F}$ & $f_{\text M}$ & $f_{\text F}$ & $f_{\text M}$ & $f_{\text F}$ \\
\mr
Edinburgh & 161 & 116 & 45 & 0.30 & 0.13 & 0.24& 0.20 & 0.22 & 0.22& 0.23 & 0.44\\
Hull & 46 & 40 & 6 & 0.20 & 0.00 & 0.33 & 0.00 & 0.25 & 0.50 & 0.23 & 0.50 \\
Manchester & 258 & 198 & 60 & 0.21 & 0.03 & 0.30 & 0.20 & 0.29 & 0.27 & 0.21 & 0.50 \\
\br
\end{tabular}
\end{indent}
\end{table}

This illustrates that the fraction of male
students in each of the four quartiles is approximately equal, and
furthermore this is consistent for the male student cohorts from all
three institutions. In other words, prior to instruction, male students are
distributed approximately evenly throughout the ability range as
determined by the FCI. In contrast, there is a starkly different
picture for female students. Across all three institutions,
approximately half the female students in each of the institutions
 are in the lowest ability quartile prior to instruction (final
column in Table~\ref{tab:quartiles}). Taken together,
Fig~\ref{fig:quartiles} and Table~\ref{tab:quartiles} present a
worrying picture, for both the starting point and outcome for female
students. Approximately half start in the lowest quartile, the majority remain there, 
and for
these students, Fig~\ref{fig:quartiles} shows that their
post-instruction test performance remains, on average, the lowest of
all 8 sub-cohorts for the larger data sets from Edinburgh and
Manchester.

It may be tempting to suggest that this is due to these students
simply being weaker at the point of entry. We find no evidence for
this in their prior qualifications, though it is difficult to obtain
good, discriminating, quantitative data since most entrants arrive with very similar school
leaving qualifications, frequently at or close to the highest grade bands. Furthermore, we have far from complete data on the number of mechanics modules taken by these students during their final secondary school study of physics and maths, so even comparing students with the same grades may not reflect their prior exposure of to Newtonian mechanics. By Looking across a wider population of UK students, rather than simply the fraction that attend our institutions, there is clear evidence that female students outperform male students in school-leaving examinations, including physics \cite{RDFYP3}. In the US, the situation seems to be slightly different. Sadler \emph{et al}, considering the impact of high school and other affective experiences \cite{sadler1} have reported that the overall background of female students entering college physics was stronger in most subjects, but not in physics. Nevertheless, they conclude that Ôthe stronger academic background of females entering college physics did not appear to help them perform better than males; in fact, they performed worse than their male counterparts with the same academic backgroundsÕ.

\subsection{Item analysis}

Given these differences, it is reasonable to question whether they arise from a greater fraction of male students getting certain questions (items) on the FCI assessment correct, or whether the origin is a consistent outperformance across the entire test.

\begin{figure*}
\begin{center}
\includegraphics[width=3.0 in]{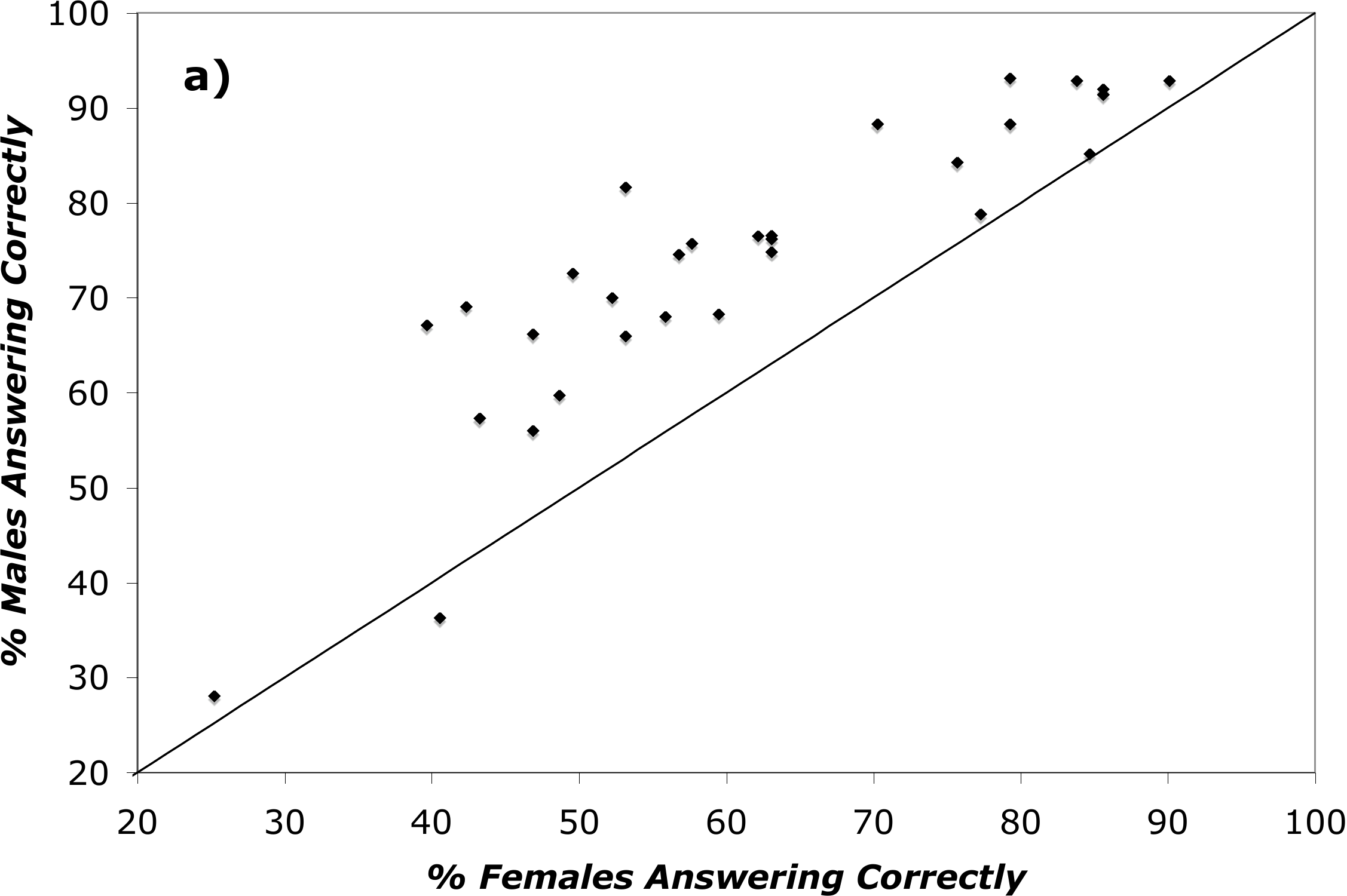}
\includegraphics[width=3.0 in]{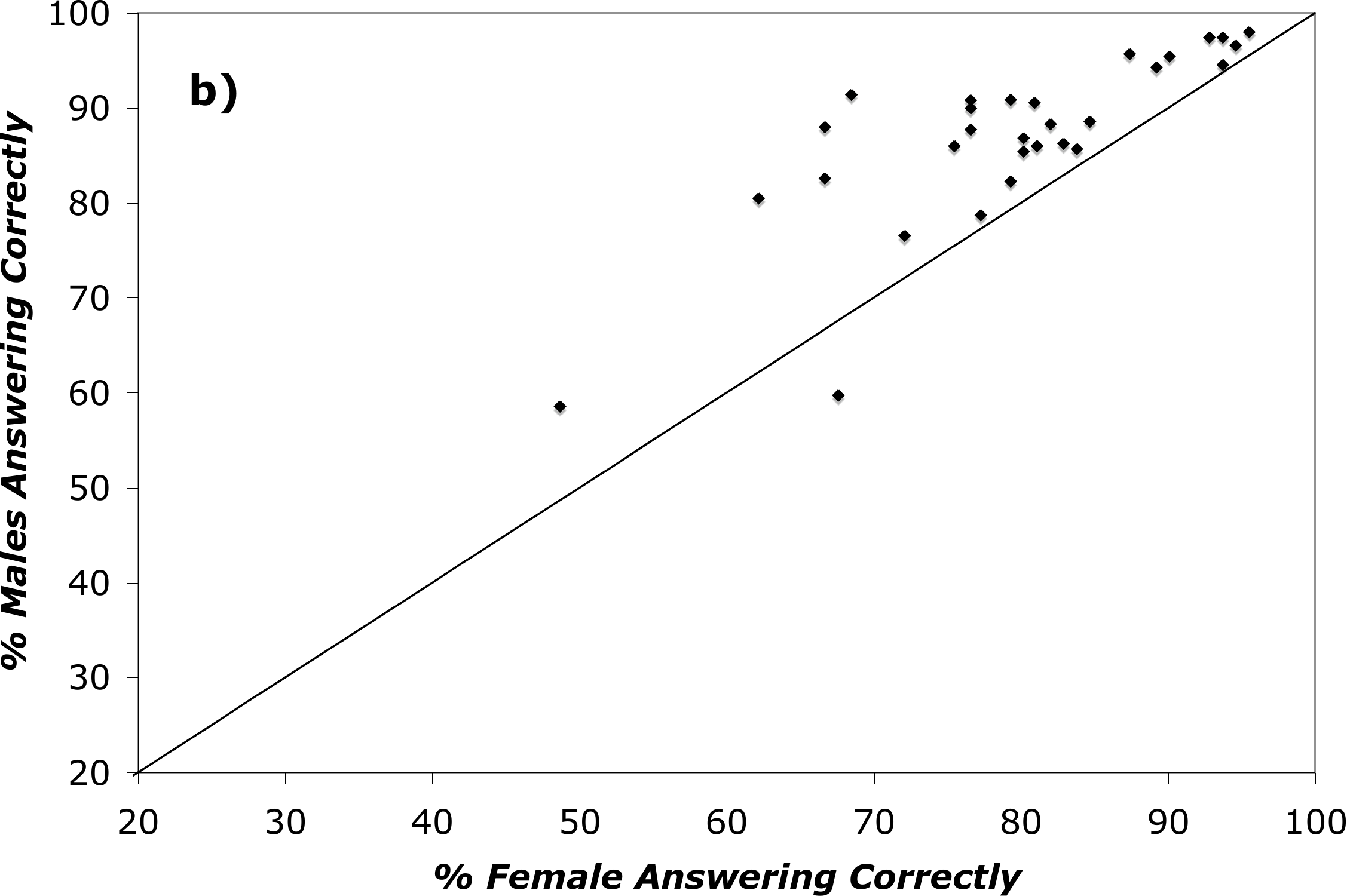}
\end{center}
\caption{\label{fig:itemsMF}Fraction of male versus fraction of female students who answer a given item on the FCI correctly. The dataset combines Edinburgh, Manchester and Hull data, and  plots (a) and (b) refer to pre- and post-instruction assessments, respectively.}
\end{figure*}

In Figure ~\ref{fig:itemsMF}, we plot the fraction of male students getting an individual item correct against the corresponding fraction of female students who do likewise, for each of the 30 items on the FCI instrument. Even though we already know we have a performance gender gap, and thus are not expecting the line of unit slope to represent a line of best fit to the data, the data confirm that a larger fraction of male students get a given item correct compared to their female counterparts, for almost every item on the instrument (i.e. the majority of points lie above the line of unit slope, with the exception of only a small number of items for both pre- and post-instruction assessments).

The largest gender differences tend to occur for items that are generally more poorly answered by the entire cohort: this is particularly evident on the pre-instruction assessment scores (left panel of Figure~\ref{fig:itemsMF}) where the spread of total scores is wider. Although a complete item-by-item analysis of differences in male / female response choices is beyond the scope of this paper, we do highlight a few illustrative examples, chosen by considering those items that lie furthest from the unit line in Figure~\ref{fig:itemsMF}, i.e. those items with the largest gender gap pre-instruction or post (or indeed both). These are in no way intended to be comprehensive, but rather representative of the complexity of the data. 

\subsubsection{Item 2} on the instrument is a companion question to
the very first question, both of which are descriptive (i.e. no
diagrams or figures presented in either the question or the possible
answer choices). In the first item, two balls are dropped from the
same height at the same time, with one being twice as heavy as the
others. Respondents are asked to choose from several options for the
relative time it will take the two balls to hit the ground. Item 2
uses the same two balls, this time rolling off a horizontal table at
the same speed. Students are asked to choose from 5 statements
describing how far away from the table the balls hit the ground. In
both cases, the correct response is that the time (in the first case)
and the distance (in the second) are the same, with the common
principle in both questions being that objects of different mass fall
at the same rate.

\begin{table}
\caption{\label{fig:item2} Proportions of male and female students correctly answering particular items on the pre- and post-instruction tests.}
\begin{center}
\begin{tabular}{llccccccccccc}
\br
Item  & Institution & \multicolumn{2}{c}{\textrm{Pre Score (\%)}} & \multicolumn{2}{c}{\textrm{Post Score (\%)}}\\ 
& & Males & Females & Males & Females\\
\mr
1 &Edinburgh & 85 & 78 & 93 & 89 \\
& Hull & 81 & 50 & 97 & 100\\
& Manchester & 92 & 83 & 95 & 97 \\
\mr 
2 &Edinburgh & 65 & 40 & 81 & 62 \\
&Hull & 58 & 33 & 75 & 17\\
& Manchester & 70 & 40 & 85 & 75 \\
\mr
13 & Edinburgh & 54 & 31 & 91 & 80 \\
& Hull & 61 & 33 & 69 & 50\\
& Manchester & 80 & 52 & 93 & 77 \\
\mr
23 & Edinburgh & 78 & 40 & 85 & 64 \\
& Hull & 75 & 50 & 86& 50\\
& Manchester & 85 & 63 & 96 & 73 \\
\br
\end{tabular}
\end{center}
\end{table}

Despite the similarity in these pair of questions, the response profiles from male and female cohorts at each of the three institutions illustrate consistent and puzzling differences for pre- and post-instruction responses, as shown in Table~\ref{fig:item2}. For item 1, there is a slightly higher fraction of male students who initially get the question correct (significantly so for Hull students, but with the caveat of small number statistics). Post-instruction there is significant improvement, and effectively no difference in the fraction of male and female students who get the question correct (which we subsequently denote as `the item gender gap'). For item 2, which is initially answered far less well by students at all three institutions, the pre-instruction item gender gap is evident and persists post instruction. Furthermore, there is no obvious incorrect answer choice chosen preferentially over others. Such response profiles - consistent across institutions, yet distinct between two linked questions -- are puzzling and merit further investigation via qualitative study.

\subsubsection{Item 13} is a descriptive question that asks students to consider the forces acting on a ball after it is thrown vertically upwards from someone's hand. The correct answer is that the only force acting on the object after it has left the thrower's hand (in the absence of air resistance) is the force of gravity alone. Well-documented student beliefs about this scenario are that the force of the `throw' persists (either as a constant or steadily decreasing force) even after the object has left the thrower's hand.

As shown in Table~\ref{fig:item2}, this item also exhibits a noticeable item gender gap, for both pre- and post-instruction tests (though much diminished in the post-instruction test). Here, the dominant incorrect choice is that the force of the throw gradually `runs out' during the upward motion, a belief still held by approximately one quarter of female students post-instruction. 

\subsubsection{Item 23} 
forms part of a set of 4 consecutive items and is a very visual
scenario, with schematic diagrams in both the question stem and the
answer choices. It is a representative example of a number of items on
the instrument: it asks students to consider the effect of (the
removal of) a constant force acting at right angles to the initial
motion. These sorts of questions, combining uniform motion in one
direction with an accelerating force applied in one perpendicular,
tend to cause students a significant challenge.

Item 23 in particular exhibits a large item gender gap pre-instruction
(Table~\ref{fig:item2}) with only marginal improvement by both male
and female cohorts post-instruction, thus resulting in a significant
post-instruction item gender gap. The pattern of post-instruction
incorrect responses of female students on this item shows a spread
across the range of all four possible incorrect options.

These examples, together with the proportion of male and female students getting other items correct, and the resulting answer choices distributions, 
illustrate a complex picture, with no obvious or immediate general patterns of behaviour between male and female cohorts. We intend to try and unravel some of these issues in an on-going qualitative study that will use some of these items (or equivalent / isomorphic questions) as the basis for discussion in structured interview scenarios. 

\subsection{Exam performance} 
Of course, the FCI covers only part of the course material. Looking
more widely, we may wish to consider the existence, or otherwise, of
performance gender gaps on final examinations (typically the principal
assessment component for all of these courses). In doing so, we accept
the inherent limitations that exams often test a degree of knowledge
(`bookwork') and other measures of proficiency as well as the conceptual
understanding that forms the focus of the FCI. Nevertheless, for the
 same cohorts of students at each of the 3 institutions in this
study, we find that there is no statistically significant difference
in examination performance for male and female cohorts, as determined by 
a $t$-test  at the 5\% significance level.

\section{Discussion}

Our results present a picture of a 
persistent and very noticeable gender gap in
both pre- and post-instruction FCI performance. This is true across all three
institutions, and appears to be independent of the precise details of method of research-informed instructional delivery. 
Data
from the University of Manchester over the past 5 years shows a similarly consistent picture over time \cite{BirchWaletAnalysis}. 
The presence of such a gap pre-instruction is entirely consistent with previous work reported from the US context \cite{CrouchMazur, DocktorHeller, Noah, fink2}, and the 
persistence of it post-instruction consistent with some \cite{Noah, fink2} but not other reports \cite{CrouchMazur}

A comparison with the Minnesota data \cite{DocktorHeller} suggests that many of the same
items have a gender difference for students from both the UK and USA. Since we can
extract only the fraction of correct answers from their paper, we can
not judge whether the same distractors have been chosen in both cases. 
It would be quite interesting to
have some data from a country with a very different educational
system, to study the effect of cultural differences on the gender
difference we have noted. The only published work we are aware of is
on Turkish high school students \cite{Turkey}, and shows little gender
difference, but at a very early stage of education and concomitant low
level of achievement, making it less relevant to the
current discussion.

Our analysis raises the question whether the FCI
instrument is partially to blame for these difficulties. There has at
least been one attempt to make a less male-stereotyped version of the FCI \cite{mccullough},
though  results using this were largely inconclusive due to a low overall attainment both pre- and post-instruction on the refined instrument, obscuring any potential real effect. Moreover we see 
substantial differences in our results on questions that do not have gender-stereotyped contexts. 


Potentially the most significant result in this study is the finding
that the lowest performing quartile on the basis of the
pre-instruction test comprises approximately half the total number of
female students in a given cohort (for Manchester and Edinburgh
data). Furthermore, the majority of these female students (70\% for
Edinburgh and 60\% for Manchester data) remain there in the
post-instruction assessment. This should serve as a warning:
instruction that otherwise looks highly effective -- as judged by
peers, pass rates and even `headline' measures of effectiveness such
as cohort-averaged improvement between pre- and post-instruction
testing -- may be far from a panacea for some of the particular 
sub-groups one would like to target especially.

Notwithstanding the persistence of the gender gap on the FCI
instrument, we believe that the outlook over the longer period of a
student's studies is not quite so bleak.  Newtonian mechanics is only
part of the first year of study, and there is some evidence to suggest
that there is a more equal distribution between quartiles based on
exam marks, especially when we look later in their programmes of
study. In a typical US programme \cite{Noah, fink2} it has been argued
that the gender effects increase. In the UK, female students who
persist to the end of the undergraduate course tend to do well, but
there is concern about the loss of students on the way there.

\section{Conclusions}

We have found a significant and persistent performance gender gap within three separate student cohorts in introductory physics at three different institutions. The gap is reduced, but not eliminated after instruction, which is in line with some (but not all) research findings from the equivalent US context. 
We find that approximately half the female students in a particular class are initially in the lowest performance quartile (on the basis of pre-instruction assessment with the FCI) and that the majority of these students remain there at the time of post-instruction testing with the same instrument. Analysis of individual items shows that male students outperform female students on practically all items on the instrument, both pre- and post-instruction. Looking at other assessments taken by these students in their respective courses, specifically examinations at the end of the course, we find no significant gender gap in performance at any of the three institutions in this study.

This study opens up several interesting avenues for further work. Clearly having identified a potential problem,
we need to find out more about causes and potential remedies. Our
first step is to follow up with a qualitative approach using
structured interviews to identify some of the factors and issues that
make a difference in the students' performance.  At the same time we
intend to continue to make use of the FCI to monitor performance in this small area of the curriculum, and study the effects
of specific interventions on the performance of male and female students.

\end{document}